\documentclass[unicode,runningheads,envcountsame,envcountsect,a4paper,dvipsnames]{llncs}
\usepackage[T1]{fontenc}
\usepackage{orcidlink}
\renewcommand{\orcidID}[1]{\orcidlink{#1}}

\usepackage{bbding}

\usepackage{graphicx}
\usepackage{subcaption}
\usepackage{xspace}
\usepackage{complexity}
\usepackage{tcolorbox}
\usepackage{amsmath}
\usepackage{amssymb}
\usepackage{cleveref}
\usepackage{mathtools}
\usepackage{tikz}
\usetikzlibrary{positioning,arrows.meta,fit,calc}

\spnewtheorem{claim}{Claim}[theorem]{\itshape}{\upshape}
\Crefname{claim}{Claim}{Claims}

\newcommand{\bondage}{\textsc{Bondage}\xspace}
\newcommand{\onebondage}{1\textsc{-Bondage}\xspace}
\newcommand{\dbondage}{$d$\textsc{-Bondage}\xspace}
\newcommand{\SigmaTwoP}{\ensuremath{\ComplexityFont{\Sigma}_\ComplexityFont{2}^{\P}}\xspace}
\newcommand{\ThetaTwoP}{\ensuremath{\ComplexityFont{\Theta}_\ComplexityFont{2}^{\P}}\xspace}
\newcommand{\DeltaTwoP}{\ensuremath{\ComplexityFont{\Delta}_\ComplexityFont{2}^{\P}}\xspace}
\newcommand{\PiTwoP}{\ensuremath{\ComplexityFont{\Pi}_\ComplexityFont{2}^{\P}}\xspace}
\newcommand{\paraNP}{\ComplexityFont{para}-\NP}

\newcommand{\tw}{\operatorname{tw}}
\newcommand{\cvd}{\operatorname{cvd}}
\newcommand{\nd}{\operatorname{nd}}

\let\doendproof\endproof
\renewcommand\endproof{~\hfill$\qed$\doendproof}

\begin{document}
\title{Computational Complexity \\ of the Bondage Problem\thanks{%
Partially supported by JSPS KAKENHI Grant Numbers 
JP22H00513, 
JP26K23808, 
JP24K02898, 
JP25K03077, 
by JST CRONOS, Grant Number JPMJCS24K2. 
}}
%
%
\author{
Shunsuke Hamada\inst{1}\orcidID{0009-0006-8928-2594}\textsuperscript{\Envelope} \and
Hiraku Morimoto\inst{1}\orcidID{0009-0002-4646-4102} \and
Hirotaka Ono\inst{1}\orcidID{0000-0003-0845-3947}
}
\authorrunning{S. Hamada et al.}
%
\institute{
Nagoya University, Nagoya, Japan.
\email{hamada.shunsuke.z9@s.mail.nagoya-u.ac.jp},
\email{hiraku.morimoto@nagoya-u.jp},
\email{ono@nagoya-u.jp}
}
\maketitle              
\begin{abstract}
For a graph $G$, the domination number $\gamma(G)$ is the minimum size of a vertex set that dominates all vertices of $G$.
The bondage number $b(G)$ is the minimum number of edges whose deletion increases the domination number.
The $d$-\bondage problem asks, given a graph $G$ and an integer $d$, whether $b(G)\leq d$. 
The precise computational complexity of \bondage has remained open. 
Most recently, Bakal and Borse have shown that even $1$-\bondage is both \NP-hard and \coNP-hard.
We show that, for every fixed integer $d\geq 1$, $d$-\bondage is $\ThetaTwoP$-complete even on chordal graphs, and that the general \bondage problem is $\SigmaTwoP$-complete.
We then study the parameterized complexity of \bondage. 
On the positive side, we give fixed-parameter algorithms parameterized by the
treewidth of the input graph alone, by the cluster vertex deletion
number plus $d$, and by the neighborhood diversity plus $d$. 
In contrast, parameterization by the domination number does not lead to tractability: \bondage is \paraNP-hard parameterized by $\gamma(G)$, and remains \W[1]-hard parameterized by $\gamma(G)$ even when $d=1$.
More precisely, the problem is polynomial-time solvable when $\gamma(G)=1$, but becomes \NP-hard already on split graphs with $\gamma(G)=2$.
These results determine the classical complexity of \bondage 
within the polynomial hierarchy and give a broad picture of its
parameterized complexity under natural graph parameters.
\end{abstract}

\section{Introduction}
A dominating set of a graph $G$ is a set of vertices such that every vertex outside the set has a neighbor in it. The minimum size of a dominating set is the domination number $\gamma(G)$, a well-studied parameter in graph theory. A natural question is how stable this optimum is under modifications of the graph. In this direction, Fink et al.~\cite{DBLP:journals/dm/FinkJKR90} introduced the \emph{bondage number} $b(G)$, defined as the minimum number of edges whose deletion increases the domination number. The corresponding \bondage problem asks, given a graph $G$ and an integer $d$, whether $b(G)\le d$. For a fixed integer $d$, we call this problem \dbondage.

Despite its long history, the precise computational complexity of \bondage has remained unresolved.  
Hu and Xu~\cite{DBLP:journals/jc/HuX12} proved that \bondage is \NP-hard and conjectured that it does not belong to \NP.
Recently, Bakal and Borse~\cite{DBLP:conf/iwoca/BakalB26} showed that even \textsc{1-Bondage} is both NP-hard and coNP-hard, thereby confirming this conjecture under the standard assumption $\NP \neq \coNP$. 
These results demonstrate that \bondage has complexity beyond the usual NP framework, but they do not determine its precise location in the polynomial hierarchy.

In this paper, we first settle this question. We show that, for every fixed
integer $d \ge 1$, $d$-\bondage is $\ThetaTwoP$-complete even on
chordal graphs, where $\ThetaTwoP=P^{\mathrm{NP}[O(\log n)]}$.
Moreover, when the deletion bound is part of the input, we prove that
\bondage is $\SigmaTwoP$-complete. Thus, the fixed-bound and
general versions are complete for $\ThetaTwoP$ and $\SigmaTwoP$,
respectively, giving a precise classification of the classical
\bondage problem.

We next investigate tractability under structural restrictions from the
viewpoint of parameterized complexity.
We obtain fixed-parameter tractability under
several structural parameters, including treewidth. In contrast,
parameterization by the domination number $\gamma(G)$ is \paraNP-hard,
thereby resolving the \W[1]-hardness conjecture of Bakal and Borse~\cite{DBLP:conf/iwoca/BakalB26}
in a stronger form. We further show that W[1]-hardness persists even when
the deletion bound is fixed to $d=1$. Figure~\ref{fig:overview} and
Section~\ref{sec:ourcontribution} give a detailed overview of our results.
Proofs of results marked with $[\star]$ are deferred to the full version.

\subsection{Related work}
Domination is a classical and extensively studied topic in graph theory, and the
minimum dominating set problem is one of the standard \NP-hard graph
optimization problems.  Numerous variants of domination and their
algorithmic and structural properties have been studied extensively~\cite{DBLP:books/daglib/0093827}.

The bondage number was introduced by Fink et al.~\cite{DBLP:journals/dm/FinkJKR90}
in 1990 as a measure of the vulnerability of domination under edge
failures. Their original work established bounds on the bondage number
and determined its exact value for several basic graph classes, including
complete graphs, paths, cycles, and complete multipartite graphs.
Since then, the bondage number has been studied extensively from a
graph-theoretic viewpoint, with particular attention to upper and lower
bounds, exact values for special graph classes, planar graphs, and various
generalizations.  We refer to the survey of Xu~\cite{xu2013bondage} for an
overview of these developments.

From the algorithmic viewpoint, the picture has been much less complete.
A linear-time algorithm is known for trees, whereas Hu and Xu~\cite{DBLP:journals/jc/HuX12}
show that \bondage is NP-hard on general graphs and conjecture that it
does not belong to NP. NP-hardness is also known for bipartite graphs~\cite{DBLP:journals/tcs/HuS14}.
Bouquet~\cite{DBLP:journals/corr/abs-2107-11216} showed that even $1$-\bondage is NP-hard on
3-regular planar graphs. Together with the general bound
$b(G)\le 2\overline d(G)-1$~\cite{DBLP:journals/dm/HartnellR99}, this immediately implies \paraNP-hardness
when parameterized by the bondage number $b(G)$ itself.
Bakal and Borse~\cite{DBLP:conf/iwoca/BakalB26} have shown that even
1-\bondage is both \NP-hard and \coNP-hard, already on split graphs.

Bakal and Borse~\cite{DBLP:conf/iwoca/BakalB26} also study the
parameterized complexity of \bondage.  Since it is \NP-hard already
for $d=1$, parameterization by the solution size $d$ alone does not even
yield an \XP{} algorithm unless $\P=\NP$.  On the positive side, for every
fixed $d$, they show that \dbondage is fixed-parameter tractable when
parameterized by clique-width.  Their brute-force argument further gives
an \XP{} algorithm for the combined parameter $d+\operatorname{cw}(G)$,
and they leave open whether this combined parameter yields an FPT
algorithm.  They also conjecture that \bondage is \W[1]-hard when
parameterized by the domination number $\gamma(G)$.  

Parameterized algorithms for domination problems often use structural
parameters such as treewidth. For \bondage, however, standard techniques
for \textsc{Dominating Set} do not directly imply tractability.
The problem asks whether there exists an edge set $F$ such that
$\gamma(G-F)>\gamma(G)$, combining a graph modification with a
comparison of two optimum values. In particular, it does not appear
to follow directly from standard MSO-based meta-theorems for
bounded-treewidth and bounded-clique-width
graphs~\cite{DBLP:journals/iandc/Courcelle90,DBLP:journals/mst/CourcelleMR00}.
This motivates a study of the parameterized complexity of
\bondage under natural structural parameters.

\bondage is closely related to blocker and interdiction problems.
Mahdavi Pajouh et al.~\cite{DBLP:journals/eor/PajouhWBP15} study the minimum edge
blocker dominating set problem, which seeks a minimum set of edges
whose deletion raises the domination value to a prescribed threshold.
More generally, blocker problems for graph parameters have been studied
extensively~\cite{DBLP:journals/gc/BazganBPR15}. Very recently, Gr{\"u}ne and Wulf~\cite{DBLP:journals/tcs/GruneW26} prove
$\SigmaTwoP$-completeness for a minimum-cardinality
interdiction version of \textsc{Dominating Set}, where selected vertices are forbidden in solutions, rather than deleted from the graph.
As they note, these two notions do not coincide in general. 
\subsection{Our contributions}\label{sec:ourcontribution}
\begin{figure}[t]
\centering
\begin{minipage}[t]{0.48\textwidth}
\centering
\resizebox{\linewidth}{!}{%
\begin{tikzpicture}[
  >=Latex,
  class/.style={
    draw,
    rounded corners,
    minimum width=22mm,
    minimum height=8mm,
    align=center
  },
  result/.style={
    draw,
    rounded corners,
    thick,
    fill=blue!10,
    minimum width=31mm,
    minimum height=10mm,
    align=center
  },
  every node/.style={font=\small}
]

\node[class] (p) at (0,0) {$\P$};

\node[class] (np)   at (-2.6,1.5) {\NP};
\node[class] (conp) at ( 2.6,1.5) {\coNP};

\node[result] (theta) at ($(np)!0.5!(conp)+(0,1.4)$)
  {$\ThetaTwoP$\\[-1mm]
   {\scriptsize fixed $d$: complete}};

\node[class] (delta) at ($(theta)+(0,1.35)$)
  {$\DeltaTwoP=\P^{\NP}$};

\node[result] (sigma) at ($(np)+(0,4.2)$)
  {$\SigmaTwoP$\\[-1mm]
   {\scriptsize general $d$: complete}};

\node[class] (pi) at ($(conp)+(0,4.2)$)
  {$\PiTwoP$};

\draw[->] (p) -- (np);
\draw[->] (p) -- (conp);

\draw[->] (np) -- (theta);
\draw[->] (conp) -- (theta);
\draw[->] (np) -- (sigma);
\draw[->] (conp) -- (pi);

\draw[->] (theta) -- (delta);

\draw[->] (delta) -- (sigma);
\draw[->] (delta) -- (pi);

\node[
  draw,
  dashed,
  rounded corners,
  fit=(theta)(delta)(sigma)(pi),
  inner xsep=3mm,
  inner ysep=2mm,
  label=above:{\small Second level of the polynomial hierarchy}
] {};
\end{tikzpicture}
}
\caption*{(a) Classical complexity}
\end{minipage}
\hfill
\begin{minipage}[t]{0.48\textwidth}
\centering
\resizebox{\linewidth}{!}{%
\begin{tikzpicture}[
  >=Latex,
   every node/.style={font=\Large},
   par/.style={
    draw,
    rounded corners,
    minimum width=23mm,
    minimum height=7mm,
    align=center
  },
  aux/.style={
    draw,
    rounded corners,
    dashed,
    minimum width=20mm,
    minimum height=7mm,
    align=center
  },
  ours/.style={
  draw=blue!70!black,
  very thick,
  fill=blue!8
  },
  rel/.style={->, thick},
  strata/.style={gray!55}
]

\draw[strata] (-3.6,6.5) -- (3.6,6.5);
\draw[strata] (-3.6,4.00) -- (3.6,4.00);
\draw[strata] (-3.6,2.05) -- (3.6,2.05);

\node[anchor=west,font=\Large\bfseries] at (-6.7,7.50)
  {\paraNP-hard};
\node[anchor=west,font=\Large\bfseries] at (-6.7,5.00)
  {\W[1]-hard};
\node[anchor=west,font=\Large\bfseries] at (-6.7,3.00)
  {\XP};
\node[anchor=west,font=\Large\bfseries] at (-6.7,1.00)
  {\FPT};

\node[par] (d)     at (-1.5,9.10) {$d$};
\node[par,ours] (gamma) at ( 1.0,9.10) {$\gamma(G)$};
\node[par] (b) at (2.9,7.3) {$b(G)$};

\node[
  right,
  font=\normalsize,
  align=left,
  draw=blue!70!black,
  densely dashed,
  rounded corners,
  fill=blue!8,
  inner xsep=4pt,
  inner ysep=2pt
]
at (0.9,8.0)
{$b(G)\le 4\,\operatorname{tw}(G)-2$};

\node[par,ours] (gammad) at (0.0,5.00) {$\gamma(G)+d$};

\node[font=\normalsize,anchor=west]
  at (-1.5,4.40) {hard even for $d=1$};

\node[par] (cwd) at (-1.5,3.00)
  {$\operatorname{cw}(G)+d$};

\node[par,ours] (nd)  at (-2.7,1.00)
  {$\operatorname{nd}(G)+d$};

\node[par,ours] (cvd) at ( 0.0,1.00)
  {$\operatorname{cvd}(G)+d$};

\node[par,ours] (tw)  at ( 2.9,1.00)
  {$\operatorname{tw}(G)$};

\draw[rel] (gammad) -- (gamma);
\draw[rel] (gammad) -- (d);

\draw[rel] (cwd) -- (d);
\draw[rel] (nd) -- (cwd);
\draw[rel] (cvd) -- (cwd);


\draw[->, thick, blue!70!black] (tw) -- (b);

\end{tikzpicture}
}
\caption*{(b) Parameterized complexity}
\end{minipage}

\caption{Overview of our results. Results established in this paper are
highlighted; dashed highlighting denotes a structural relation.}
\label{fig:overview}
\end{figure}

Figure~\ref{fig:overview} summarizes the complexity landscape of
\bondage, including both previously known results and the results
established in this paper. 
Our main contributions are as follows. 
\begin{itemize}
\item
We determine the classical complexity of \bondage within the second
level of the polynomial hierarchy. For every fixed integer $d\ge 1$,
$d$-\bondage is $\ThetaTwoP$-complete, even on chordal graphs.
When $d$ is part of the input, \bondage is $\SigmaTwoP$-complete. 
Thus, the long-standing question of whether \bondage lies beyond NP is resolved up to standard complexity assumptions. (Section~\ref{sec:ph})

\item
On the tractability side, we develop a treewidth DP that tracks
the increase of partial domination optima under edge deletions, yielding
FPT for $\tw(G)+d$. We further prove the structural bound
$b(G)\le 4\tw(G)-2$ for graphs without isolated vertices, which yields
FPT parameterized by $\tw(G)$ alone. We also obtain \FPT{} algorithms for
$\nd(G)+d$ and $\cvd(G)+d$. 
(Section~\ref{sec:fpt})

\item
On the intractability side, we show that parameterization by the
domination number $\gamma(G)$ does not yield tractability, though \bondage is polynomial-time solvable when
$\gamma(G)=1$.
More precisely, we show that \bondage is \paraNP-hard when parameterized by
$\gamma(G)$, already for split graphs with $\gamma(G)=2$,
thereby settling the \W[1]-hardness conjecture of Bakal and Borse~\cite{DBLP:conf/iwoca/BakalB26} 
in a stronger form. 
We further show that \W[1]-hardness persists
even when the deletion bound is fixed to $d=1$. (Section \ref{sec:parahard})
\end{itemize}

\section{Preliminaries}

We assume that the reader is familiar with standard notions from polynomial hierarchy and parameterized complexity (see e.g. \cite{DBLP:books/daglib/0023084,DBLP:series/txtcs/FlumG06}).
We follow the standard notations in graph theory (see e.g. \cite{diestel2025graph}).

Let $G = (V, E)$ be a graph.
For $v \in V$, let $N(v)$ and $N[v] = N(v) \cup \{v\}$ denote its open and closed neighborhoods, respectively.
Two distinct vertices $u,v\in V$ are \emph{twins} if $N_G(u)\setminus\{v\}=N_G(v)\setminus\{u\}$.
They are \emph{true twins} if they are adjacent; equivalently,
$N_G[u]=N_G[v]$.
Let $\overline{d}(G)$ denote its average degree.
A set $D\subseteq V$ is a \emph{dominating set} of $G$ if every vertex
in $V\setminus D$ has a neighbor in $D$.
The \emph{domination number} of $G$, denoted by $\gamma(G)$, is the
minimum cardinality of a dominating set of $G$.
A set $S \subseteq V$ is a \emph{vertex cover} of $G$ if, for every edge $e \in E$, we have $e \cap S \neq \emptyset$.
The \emph{vertex cover number} of $G$, denoted by $\tau(G)$, is the
minimum cardinality of a vertex cover set of $G$.
For $X \subseteq V$ and $F \subseteq E$, let $G-X$ and $G-F$ denote the graphs obtained by deleting $X$ and $F$, respectively.
The \emph{bondage number} of $G$, denoted by $b(G)$, is the minimum
cardinality of $F\subseteq E$ such that $\gamma(G-F)>\gamma(G)$.
\bondage asks, given a graph $G$ and an integer $d \geq 0$, whether $b(G)\le d$.
For a fixed integer $d$, we call this problem \dbondage.

Throughout this paper, all graphs are finite, simple, and undirected.
We may assume without loss of generality that the input graph $G$ is connected.
We further assume that $G$ has at least two vertices.

\section{Classical Complexity in the Polynomial Hierarchy}\label{sec:ph}
In this section, we determine the complexity of \bondage within the
second level of the polynomial hierarchy.  We first consider the case
where the deletion bound $d$ is fixed, and then turn to the general
case where $d$ is part of the input.

\subsection{Completeness for \ThetaTwoP}

We begin with the case where the deletion bound $d$ is fixed. 
The computational complexity class \ThetaTwoP, also denoted by $\P^{\NP[\mathcal{O}(\log{n})]}$, is the class of languages decidable by a deterministic polynomial-time Turing machine that makes $\mathcal{O}(\log{n})$ queries to an {\NP} oracle~\cite{DBLP:journals/siamcomp/Wagner90}.

\begin{theorem}\label{thm:theta2p-membership}
  For every fixed integer $d \geq 1$, \dbondage belongs to \ThetaTwoP.
\end{theorem}
\begin{proof}
    For fixed integer $d \geq 1$, given a graph $G$ and an integer $\ell$, consider the problem of deciding whether $\gamma(G-F) \leq \ell$ for every $F \subseteq E(G)$ with $|F| \leq d$.
    This problem is in \NP: since $d$ is fixed, there are only polynomially many such sets $F$, and a certificate consists of a dominating set of size at most $\ell$ for each $G-F$.
    We compute $\gamma(G)$ by binary search using $\mathcal{O}(\log{|V(G)|})$ \NP-oracle queries and then make one additional query to the above problem with $\ell = \gamma(G)$.
    Since deleting edges cannot decrease the domination number, $b(G) \leq d$ if and only if this final oracle query returns no.
    Hence \dbondage belongs to \ThetaTwoP.
\end{proof}

For a graph $G$, let $\tau\emph{-anticore}(G)$ denote the set of vertices contained in no minimum vertex cover of $G$.
We reduce from $\tau\textsc{-Anticore}$, which asks whether $|\tau\text{-anticore}(G)| \geq 1$.
A graph $G$ is $\tau\emph{-vertex-critical}$ (called $\beta\emph{-vertex-critical}$ in \cite{DBLP:journals/algorithmica/BurjonsFHKW24}) if $\tau(G - v) < \tau(G)$ for every vertex $v \in V(G)$.
Burjons, Frei, Hemaspaandra, Komm, and Wehner proved that recognizing $\tau\text{-vertex-critical}$ graphs is \ThetaTwoP-complete~\cite[Theorem 18]{DBLP:journals/algorithmica/BurjonsFHKW24}.
Observe that $\tau(G-v)<\tau(G)$ if and only if $v$ belongs to some minimum vertex cover of $G$.
Thus, $G$ is $\tau$-vertex-critical if and only if $\tau\text{-anticore}(G)=\emptyset$.
Since $\ThetaTwoP$ is closed under complementation, $\tau$-\textsc{Anticore} is $\ThetaTwoP$-complete.

\begin{theorem}\label{thm:theta2p-hard-one}
  \onebondage is \ThetaTwoP-hard even when restricted to chordal graphs.
\end{theorem}
\begin{proof}
    Given a graph $G$, let $n = |V(G)|$ and construct $H$ as follows (see \Cref{fig:theta2p:2}).
    Make $V(G)$ a clique.
    For every edge $e = \{u, v\} \in E(G)$, add two vertices $x_e, y_e$, each adjacent to $u$ and $v$.
    For every $v \in V(G)$, add vertices $a_v, b_v$ and the edges $\{v, a_v\}$ and $\{a_v, b_v\}$.
    \begin{figure}[t]
        \centering
        \includegraphics[page=1]{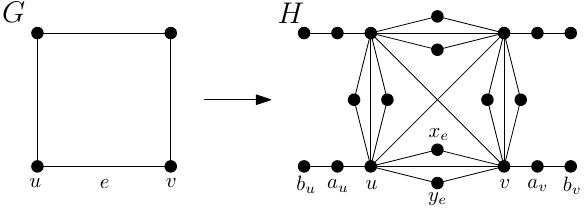}
        \caption{An example of graph $G$ and the corresponding graph $H$.}
        \label{fig:theta2p:2}
    \end{figure}

    The construction satisfies $\gamma(H) = n + \tau(G)$.
    Indeed, a minimum vertex cover $S$ gives the dominating set $S \cup \{a_v \mid v \in V(G)\}$.
    Conversely, in a minimum dominating set we may choose $a_v$ rather than $b_v$, and every selected $x_e$ or $y_e$ can be replaced by an endpoint of $e$; the remaining selected vertices therefore yield a vertex cover of $G$.
    
    We show that $\tau\text{-anticore}(G) \neq \emptyset$ if and only if $b(H) = 1$.
    If $v \in \tau\text{-anticore}(G)$, deleting $\{a_v, b_v\}$ forces every minimum dominating set of the resulting graph to induce a vertex cover containing $v$, and hence increases the domination number.
    Conversely, if the $\tau\text{-anticore}$ is empty, every vertex belongs to some minimum vertex cover in $G$; using such a cover, one can construct a dominating set of size $n + \tau(G)$ after deleting any single edge of $H$.
    
    Finally, $H$ is chordal: a perfect elimination ordering is obtained by eliminating all $x_e, y_e$, then each $b_v, a_v$, and finally the vertices of $G$.
    Thus, \onebondage is \ThetaTwoP-hard on chordal graphs.
\end{proof}

We now extend the above result to \dbondage for every fixed $d \geq 2$.

\begin{theorem}\label{thm:theta2p-hard-k}
  For every fixed $d \geq 2$, \dbondage is \ThetaTwoP-hard even when restricted to chordal graphs.
\end{theorem}
\begin{proof}
    For fixed $d \ge 2$, we extend the construction above.
    For every $v \in V(G)$, replace $v$ by a clique $C_v$ of $2d+1$ true twins, each having the former neighbors of $v$.
    Similarly, replace each $b_v$ by a clique $B_v$ of $2d-1$ true twins, each adjacent to $a_v$.
    Thus, $R_v:=B_v \cup \{a_v\}$ induces a clique of size $2d$.
    Denote the resulting graph by $H_d$.
    By construction, we have $\gamma(H_d)=\gamma(H)$, and $\tau\text{-anticore}(G)\neq\emptyset$ if and only if $b(H_d)\le d$.
%
The forward direction deletes a perfect matching of $R_v$; the reverse uses untouched twins, or a minimum vertex cover containing $v$ in the matching case.
This construction extends the above reduction to every fixed $d \ge 2$.
Moreover, replacing a vertex by a clique of true twins preserves chordality, so $H_d$ is chordal.
Hence, \dbondage is \ThetaTwoP-hard on chordal graphs.
\end{proof}


\begin{corollary}
  For every fixed integer $d \geq 1$, \dbondage is \ThetaTwoP-complete even when restricted to chordal graphs.
\end{corollary}

\subsection{Completeness for $\SigmaTwoP$}

\begin{theorem}
  \bondage is \SigmaTwoP-complete.
\end{theorem}
\begin{proof}
    Membership in \SigmaTwoP is immediate: nondeterministically guess an edge set $F$ with $|F| \leq d$ and compare $\gamma(G-F)$ and $\gamma(G)$ using an \NP-oracle.
    
    To show the hardness, we reduce from $\textsc{QSAT}_2$ for 3-DNF formulas.
    $\textsc{QSAT}_2$ asks whether there exists an assignment to $X$ such that $\Phi(X,Y)$ is true for every assignment to $Y$. 
    Let $\Phi(X, Y) = T_1 \lor \cdots \lor T_\ell$, $X = \{x_1, \dots, x_n\}$, $Y = \{y_1, \dots, y_m\}$, and let $M = 2n + 1$.
    It is known that $\textsc{QSAT}_2$ is \SigmaTwoP-complete even when restricted to 3-DNF formulas~\cite{DBLP:journals/tcs/Marx08}.
    We may assume that $n \geq 1$, and identify true and false with $1$ and $0$, respectively. 

    Construct a graph $G$ as follows (see \Cref{fig:sigma2p}).
    For each $x_i$, let $X_i = \{x_i, \bar{x}_i, a_i,$ $d_i^1, \dots, d_i^M, p_i, q_i\}$ induce a clique minus an edge $\{p_i, q_i\}$, and add $h_i$ adjacent to $x_i, \bar{x}_i$.
    For each $y_j$, let $Y_j = \{y_j, \bar{y}_j, e_j^1, \dots, e_j^M, r_j, s_j\}$ induce a clique minus an edge $\{r_j, s_j\}$.
    For each term $T_t$, add a vertex $c_t$ adjacent to the literal vertex corresponding to the complement of every literal in $T_t$.
    Finally, make every $a_i$ adjacent to all $h_1, \dots, h_n, c_1, \dots, c_\ell$, and set $d = n$.
    \begin{figure}[t]
        \centering
        \includegraphics[page=1]{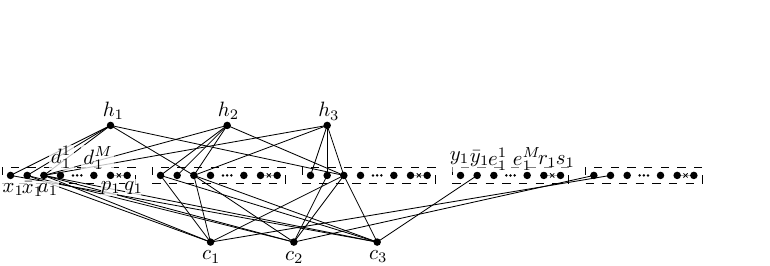}
        \caption{An example of the graph $G$ corresponding to $\Phi = (x_1 \land \bar{x}_2 \land y_2) \lor (x_1 \land x_3 \land \bar{y}_2) \lor (\bar{x}_1 \land \bar{x}_2 \land y_1)$. The vertices $\{x_i, \bar{x}_i, a_i, d_i^1, \dots, d_i^M, p_i, q_i\}$ form a clique minus the edge $\{p_i, q_i\}$, and the vertices $\{y_j, \bar{y}_j, e_j^1, \dots, e_j^M, r_j, s_j\}$ form a clique minus the edge $\{r_j, s_j\}$.}
        \label{fig:sigma2p}
    \end{figure}

    We first observe that $\gamma(G) = n + m$.
    The pairs $p_i, q_i$ and $r_j, s_j$ force at least one vertex from each $X$- and $Y$-gadget, giving the lower bound.
    Conversely, $\{a_1, x_2, \dots, x_n, y_1, \dots, y_m\}$ is a dominating set of size $n+m$.

    Suppose first that $\Phi$ is a yes-instance, and let $\varphi_X$ satisfy
    $\Phi(\varphi_X,\varphi_Y)=1$ for every $\varphi_Y$.
    For each $i$, delete $\{a_i, \bar{x}_i\}$ if $\varphi_X(x_i) = 1$, otherwise, $\{a_i, x_i\}$.
    Let $F$ be these $n$ edges. 
    If $G-F$ had a dominating set $D$ of size $n+m$, 
 exactly one vertex would be chosen from each variable gadget; the deleted edge excludes
$a_i$, while $h_i$ forces the $X_i$-vertex to be the literal consistent with $\varphi_X$.
Choose $\varphi_Y$ consistently with the selected $Y$-literals,
arbitrarily otherwise. Hence every term vertex $c_t$ must be dominated
by a selected complementary literal.
Thus every $T_t$
contains a literal evaluating to $0$, contradicting
$\Phi(\varphi_X,\varphi_Y)=1$. 
Thus $\gamma(G-F)>\gamma(G)$, and $b(G)\le n$.

    Conversely, suppose that $\Phi$ is a no-instance, and let $F\subseteq E(G)$ with $|F|\le n$. 
    If some $a_i$ is not incident with $F$, then, since $M=2n+1$, untouched dummy vertices together with $a_i$ form a dominating set of size $n+m$.
%
    Otherwise, $|F|=n$ and exactly one deleted edge is incident with each $a_i$. 
    Define $\varphi_X(x_i)=1$ if and only if $\bar x_i$ is incident
with this edge. Since $\Phi$ is a no-instance, there exists $\varphi_Y$
with $\Phi(\varphi_X,\varphi_Y)=0$.
Selecting the literal vertex evaluating to $1$ for each variable in
$X\cup Y$ gives a dominating set of size $n+m$ in $G-F$: every
$c_t$ is dominated since $T_t$ contains a literal evaluating to $0$.
Hence $\gamma(G-F)=\gamma(G)$.
\end{proof}

\section{Fixed-Parameter Tractability}\label{sec:fpt}

\subsection{Treewidth}\label{sec:tw}

We show that \bondage is fixed-parameter tractable for the treewidth of the input graph.
Our proof proceeds in two steps.
We first give a fixed-parameter algorithm for $\mathrm{tw}(G) + d$, and then show that $d$ is upper bounded by a function of $\mathrm{tw}(G)$; consequently, \bondage is fixed-parameter tractable for treewidth alone. 

Let $k=\mathrm{tw}(G)$.
We use a nice tree decomposition $T$ of width $k$, rooted at a node $r$, whose root and leaves have empty bags.
We additionally use introduce edge nodes and assume that every edge of $G$ is introduced exactly once.
For each node $t\in V(T)$, let $X_t$ denote its bag.
Let $T_t$ denote the subtree of $T$ rooted at $t$, let $V_t=\bigcup_{s\in V(T_t)} X_s$, and let $E_t$ be the set of edges introduced at nodes of $T_t$.
We write $G_t=(V_t,E_t)$.

We first recall the standard dynamic programming for \textsc{Dominating Set} on a nice tree decomposition~\cite{DBLP:books/sp/CyganFKLMPPS15}.
For each node $t$, let $\mathcal{C}_t = \{1, 0, \hat{0}\}^{X_t}$.
For $f \in \mathcal{C}_t$, a set $D \subseteq V_t$ is \emph{compatible} with ($t$, $f$) if $D \cap X_t = f^{-1}(1)$ and $D$ dominates every vertex in $V_t \setminus f^{-1}(\hat{0})$ in $G_t$.
Thus, color $1$ means that the vertex belongs to $D$, color $0$ means that it does not belong to $D$ and the vertex must already be dominated, and color $\hat{0}$ means that it is not yet required to be dominated.

Let $c[t, f]$ be the minimum size of a set compatible with ($t$, $f$), or $+\infty$ if no such set exists.
These values are computed by the standard Dominating Set DP over leaf, introduce vertex, introduce edge, forget and join nodes~\cite{DBLP:books/sp/CyganFKLMPPS15}.
In particular, $c[r, \emptyset] = \gamma(G)$.

We now extend this DP to account for edge deletions.
The following observation allows us to store only the increase of each partial optimum.

\begin{lemma}\label{lemma:diff}{\rm[$\star$]}
  Let $F \subseteq E_t$.
  If there exists a set compatible with ($t$, $f$) in $G_t-F$, then the minimum size of such a set is between $c[t,f]$ and $c[t,f]+|F|$.
\end{lemma}

By \Cref{lemma:diff}, deleting at most $d$ edges increases every finite partial optimum by at most $d$.
We therefore describe a state by a difference function $\delta: \mathcal{C}_t \to \{0,1,\dots,d,\infty\}$.
For $F \subseteq E_t$ with $|F| \leq d$, we say that $F$ \emph{realizes} a map $\delta$ if, for every $f \in \mathcal{C}_t$, the minimum size of a set compatible with ($t$, $f$) in $G_t - F$ is $c[t,f]+\delta(f)$, with $\delta(f) = +\infty$ when no compatible set exists.

For a node $t$ and a state $\delta$, let $\mathrm{DP}[t, \delta]$ be the minimum size of an edge set $F \subseteq E_t$, $|F| \leq d$, realizing $\delta$, and set it to $+\infty$ if no such set exists.
At the root, $b(G) \leq d$ if and only if there is a state $\delta$ such that $\mathrm{DP}[r, \delta] \leq d$ and $\delta(\emptyset) \ge 1$.

\paragraph{Leaf node.}
Let $t$ be a leaf node.
Since $X_t = \emptyset$ and $E_t = \emptyset$, the only possible edge deletion set is $F = \emptyset$.
Thus, the only realizable state is $\delta_0$ with $\delta_0(\emptyset) = 0$, and we set $\mathrm{DP}[t, \delta_0] = 0$.

\paragraph{Introduce vertex node.}
Let $t$ be an introduce vertex node with child $t'$, where $X_t = X_{t'} \cup \{v\}$.
Since no edge is introduced at $t$, we have $E_t = E_{t'}$.
Moreover, $v$ is isolated in $G_t$.
Hence, for every state $\delta'$ with $\mathrm{DP}[t', \delta'] \leq d$, we define $\delta(f) = +\infty$ if $f(v)=0$, $\delta'(f|_{X_{t'}})$ otherwise,  
and update $\mathrm{DP}[t,\delta] \gets \min\{\mathrm{DP}[t,\delta],\mathrm{DP}[t',\delta']\}$.

\paragraph{Introduce edge node.}
Let $t$ be an introduce edge node with child $t'$ that introduces an edge $e = \{u,v\} \notin E_{t'}$.
Then $X_t = X_{t'}$, $E_t = E_{t'} \cup \{e\}$.
We consider whether $e$ is retained or deleted.
If $e$ is retained, the transition is the same as in the standard Dominating Set DP.
For every $\delta'$ with $\mathrm{DP}[t',\delta'] \leq d$, define 
$\delta^{\text{keep}}(f) = \delta'(f_{v \to \hat{0}})$ if $(f(u),f(v)) = (1,0)$, $\delta'(f_{u \to \hat{0}})$ if $(f(u),f(v)) = (0,1)$, $\delta'(f)$ otherwise.
Since the number of deleted edges is unchanged, we update $\mathrm{DP}[t,\delta^{\text{keep}}] \gets \min\{\mathrm{DP}[t,\delta^{\text{keep}}],\mathrm{DP}[t',\delta']\}$.
If $e$ is deleted, let $F^{\text{delete}} = F' \cup \{e\}$.
Since $G_t-F^{\text{delete}} = G_{t'}-F'$, the minimum size of a set compatible with ($t$, $f$) is $c[t', f] + \delta'(f)$.
Thus, for every $\delta'$ with $DP[t', \delta'] \leq d - 1$, define $\delta_{\text{delete}}(f)$ to be $+\infty$ if $c[t,f]$, $c[t',f]$, or $\delta'(f)$ is $+\infty$, and $c[t',f]+\delta'(f)-c[t,f]$ otherwise.
We then update $\mathrm{DP}[t,\delta^{\text{delete}}] \gets \min\{\mathrm{DP}[t,\delta^{\text{delete}}],\mathrm{DP}[t',\delta']+1\}$.

\paragraph{Forget node.}
Let $t$ be a forget node with child $t'$, where $X_t = X_{t'} \setminus \{w\}$ for some $w \in X_{t'}$.
Since no edge is introduced at $t$, we have $E_t = E_{t'}$.
For every state $\delta'$ with $\mathrm{DP}[t', \delta'] \leq d$, define $\delta(f) = \min\{c[t',f_{w \to 0}]+\delta'(f_{w \to 0}), c[t',f_{w \to 1}]+\delta'(f_{w \to 1})\}-c[t,f]$, with $\delta(f) = +\infty$ if the minimum is $+\infty$ or $c[t, f]$ is $+\infty$.
We then update $\mathrm{DP}[t,\delta] \gets \min\{\mathrm{DP}[t,\delta],\mathrm{DP}[t',\delta']\}$.

\paragraph{Join node.}
Let $t$ be a join node with children $t_1$ and $t_2$.
Since $E_t = E_{t_1} \cup E_{t_2}$ and $E_{t_1} \cap E_{t_2} = \emptyset$, an edge-deletion set for $t$ decomposes uniquely into deletion sets for the two children.
We say that $(f_1, f_2)$ are \emph{consistent} with $f$ if, for every vertex $v \in X_t$ , $(f(v), f_1(v), f_2(v)) \in \{(1,1,1), (0,0,\hat{0}), (0, \hat{0},0), (\hat{0},\hat{0}, \hat{0})\}$.
For every pair of states $(\delta_1, \delta_2)$ satisfying $\mathrm{DP}[t_1, \delta_1] +\mathrm{DP}[t_2, \delta_2] \leq d$, let $\delta(f)$ be the minimum of $c[t_1,f_1]+\delta_1(f_1)+c[t_2,f_2]+\delta_2(f_2)-|f^{-1}(1)|-c[t,f]$ over all pairs $(f_1,f_2)$ consistent with $f$, and set $\delta(f)=+\infty$ if this minimum or $c[t,f]$ is $+\infty$.
We then update $\mathrm{DP}[t,\delta] \gets \min\{\mathrm{DP}[t,\delta],\mathrm{DP}[t_1,\delta_1]+\mathrm{DP}[t_2,\delta_2]\}$.

The correctness of the recursive formula can be shown by induction over the tree decomposition. 
The above argument describes our algorithm.
%
We now analyze its running time.
The nice tree decomposition may be assumed to have $\mathcal{O}(kn)$ nodes~\cite{DBLP:books/sp/CyganFKLMPPS15}, yielding the  running time $k \cdot (d+2)^{3^{\mathcal{O}(k)}} \cdot n$.
Since a nice tree decomposition of width $k$ can be computed in fixed-parameter time for $k$~\cite{DBLP:journals/siamcomp/Bodlaender96,DBLP:books/sp/CyganFKLMPPS15,DBLP:books/sp/Kloks94}, we have the following. 
\begin{lemma}
  The \bondage problem is fixed-parameter tractable when parameterized by $\mathrm{tw}(G)+d$, where $d$ is the solution size bound.
\end{lemma}

The following bound relates the bondage number to treewidth.

\begin{theorem}
  For every graph $G$ without isolated vertices, $b(G) \leq 4\mathrm{tw}(G)-2$.
\end{theorem}
\begin{proof}
  Let $G = (V,E)$ be a graph of treewidth $k = \mathrm{tw}(G)$.
  The bound $b(G) \leq 2\overline{d}(G)-1$ holds in general~\cite{DBLP:journals/dm/HartnellR99}.
  Since $G$ has treewidth $k$, we have $|E| \leq k|V|-k(k+1)/2$~\cite{DBLP:journals/siamcomp/Bodlaender96}.
  Hence $\overline{d}(G) = 2|E|/|V| \leq 2k-k(k+1)/|V| < 2k$, which implies $b(G) < 4k-1$.
  As $b(G)$ is an integer, it follows that $b(G) \leq 4k-2$.
\end{proof}

If the input bound satisfies $d \geq 4k-2$, we return \textsc{yes}; otherwise, we run the algorithm above.
In the latter case, $d$ is bounded by a function of $k$.
Combining the preceding theorems yields our main result.
\begin{theorem}
  The \bondage problem is solvable in time $2^{2^{O(\mathrm{tw}(G))}}n$.
\end{theorem}

\subsection{Neighborhood-based parameters}\label{sec:neighbor}
We focus on two neighborhood-based parameters: neighborhood diversity
and cluster vertex deletion number. We only state their definitions and
the corresponding results.
\paragraph{Neighborhood diversity.}
A \emph{twin class} of $G$ is a maximal set of pairwise twin vertices in $G$.
The \emph{neighborhood diversity} of $G$, denoted by $\mathrm{nd}(G)$, is the number of twin classes in $G$.
Since we can compute the twin classes in $G$ in polynomial time, neighborhood diversity can be computed in polynomial time.
\begin{theorem}{\rm [$\star$]}
    \bondage is fixed-parameter tractable for $\mathrm{nd}(G) + d$.
\end{theorem}

\paragraph{Cluster vertex deletion number.} 
Let $G$ be a graph.
A set $X \subseteq V(G)$ is a cluster vertex deletion set if $G-X$ is a cluster graph, namely, each component of $G-X$ is a complete graph.
The cluster vertex deletion number of $G$, denoted by $\mathrm{cvd}(G)$, is the size of a minimum cluster vertex deletion set of $G$.
While finding the minimum cluster vertex deletion set of $G$ is \NP-hard, it is fixed-parameter tractable parameterized by $\mathrm{cvd}(G)$~\cite{DBLP:journals/mst/HuffnerKMN10}.
\begin{theorem}{\rm [$\star$]}
    \bondage is fixed-parameter tractable for $\mathrm{cvd}(G) + d$.
\end{theorem}

\section{Hardness with respect to the domination number}\label{sec:parahard}
\subsection{\paraNP-hardness}
Bakal and Borse~\cite{DBLP:conf/iwoca/BakalB26} conjecture that \bondage is \W[1]-hard for $\gamma(G)$.
It is known that the problem is polynomial-time solvable for $\gamma(G)=1$~\cite[Lemma 49]{xu2013bondage}. 
Interestingly, it becomes \NP-hard already on split graphs even with
$\gamma(G)=2$. 


\begin{theorem}
    \bondage is \NP-hard even when the graph $G$ is restricted to split graphs with $\gamma(G) = 2$, and hence
\paraNP-hard parameterized by $\gamma(G)$.
\end{theorem}
\begin{proof}
    We reduce from Vertex Cover.
    Let $\langle{H,\ell}\rangle$ be an instance with $E(H) \neq \emptyset$.
    We construct an instance $\langle{G, d}\rangle$ of \bondage from $\langle{H,\ell}\rangle$ (see \Cref{fig:paranp-split}).
    For each $e\in E(H)$, introduce two vertices $x_e^1, x_e^2$, and let $X$ be the set of all these vertices.
    Add a vertex $s$ and make $X \cup \{s\}$ a clique.
    For each $v \in V(H)$, add a vertex $a_v$ adjacent to $s$ and to $x_e^i$ whenever $e$ is not incident with $v$.
    For every distinct $x, y \in X$, add a vertex $b_{x,y}$ adjacent to $s$ and all vertices of $X \setminus \{x,y\}$.
    Finally, add two vertices $p, q$, each adjacent to every vertex of $X$, and set $d = \ell$.
    The graph is clearly a split graph, with clique $X \cup \{s\}$.
    \begin{figure}[t]
        \centering
        \includegraphics[page=1]{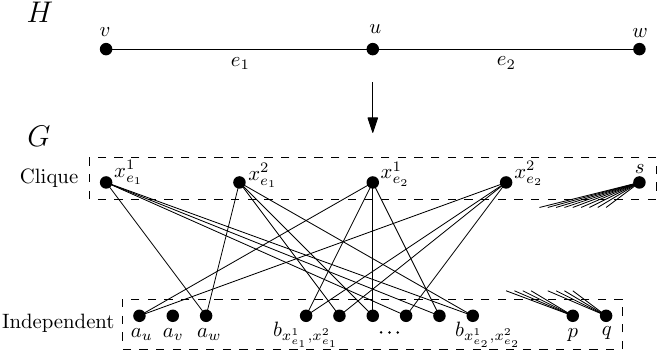}
        \caption{
        An example of the reduction from \textsc{Vertex Cover} to \bondage. The set $X\cup\{s\}$ is the clique; all other vertices are independent. Edges inside the clique are omitted.
    }\label{fig:paranp-split}
    \end{figure}

    We first observe that $\gamma(G) = 2$, and moreover the minimum dominating sets are exactly the sets $\{x,s\}$ with $x \in X$.
    Indeed, every such set dominates $G$. On the other hand, any dominating set of size two must meet $X$, since $p$ and $q$ have neighborhood $X$; and a dominating set of size two containing two vertices of $X$ misses the corresponding vertex $b_{x,y}$, while a set $\{x,w\}$ with $w \notin X \cup \{s\}$ misses one of the vertices $a_u,a_v$ associated with the endpoints of the edge corresponding to $x$.
    Thus the second vertex must be $s$.
    
    We claim that $b(G) = \tau(H)$.
    Let $C$ be a minimum vertex cover of $H$, and delete the edges $\{s,a_v\}$ for $v \in C$.
    For every minimum dominating set $\{x_e^j,s\}$, at least one endpoint $v$ of $e$ belongs to $C$, and then $a_v$ is dominated only by $s$ in $G$.
    Hence deleting $\{s,a_v\}$ destroys this dominating set.
    Therefore $\gamma(G-F)>2$, and $b(G) \le \tau(H)$.

    Conversely, let $F$ be any edge set such that $\gamma(G-F) > \gamma(G)$. 
    Define $C_F=\{v \in V(H) \mid \{s,a_v\}\in F\}$, $F' = F \setminus \{\{s,a_v\} \mid v \in V(H)\}$, and let $U$ be the set of edges of $H$ not covered by $C_F$.
    For each $e \in U$ and $j \in \{1,2\}$, the set $\{x_e^j, s\}$ is a minimum dominating set of $G$ but not of $G-F$.
    Hence some edge of $F'$ is needed to destroy it. As in the construction, such an edge can be charged to at most two vertices of $X$. Since each $e \in U$ yields two vertices $x_e^1,x_e^2$, we obtain $2|U| \le 2|F'|$, and therefore $|U| \le |F'|$.
    Choosing one endpoint from every edge in $U$, we obtain a vertex cover $C=C_F \cup \{v_e \mid e \in U\}$ of $H$ with $|C| \le |C_F|+|U|\le |C_F|+|F'|=|F|$.
    Thus $\tau(H) \le |F|$, and hence $b(G) \ge \tau(H)$. 
    Therefore $b(G) = \tau(H)$, so $\langle{H, \ell}\rangle$ is a yes-instance of Vertex Cover if and only if $\langle{G, \ell}\rangle$ is a yes-instance of Bondage.
\end{proof}


\subsection{$\W[1]$-hardness for $d=1$}
Our reduction adapts the standard reduction from \textsc{Multicolored Independent Set} to \textsc{Dominating Set} (see, e.g., \cite[Theorem 13.9]{DBLP:books/sp/CyganFKLMPPS15}).

\begin{theorem}
    \onebondage is $\W[1]$-hard parameterized by $\gamma(G)$.
\end{theorem}
\begin{proof}
    We give a parameterized reduction from \textsc{Multicolored Independent Set}, which is known to be $\W[1]$-hard for  the number $k$ of color classes~\cite{DBLP:series/mcs/DowneyF99}.
    Let $\langle{G, k, (V_1, \dots, V_k)}\rangle$ be an instance, where $V_1, \dots, V_k$ are the color classes.
    We may assume that $|V_i| \geq 2$ for every $i$:
    empty color classes can be handled directly, and a singleton class can be duplicated without changing the answer.

    We first add two vertices $a$ and $b$, where $a$ is isolated and $b$ is adjacent to every vertex of $G$, and set $V_{k+1} = \{a, b\}$.
    Let the resulting graph be $G'$.
    Clearly, this preprocessing preserves the existence of a multicolored independent set.

    
    We construct a graph $H$ from $G'$ as follows (see \Cref{fig:w1-hard}).
    First, we add all vertices of $G'$ and, for every $1 \leq i \leq k+1$, make $V_i$ a clique.
    For every edge $e = \{u, v\} \in E(G')$, where $u \in V_i$ and $v \in V_j$ with $i \neq j$, we add a new vertex $w_e$ adjacent to every vertex of $(V_i \cup V_j) \setminus \{u, v\}$.
    Next, for every $1 \leq i \leq k+1$, we add two vertices $x_i$ and $y_i$, each adjacent to every vertex of $V_i$.
    Finally, we add a vertex $z$ adjacent to every vertex of $G'$ and every vertex $w_e$.
    This completes the construction.
    \begin{figure}[t]
        \centering
        \includegraphics{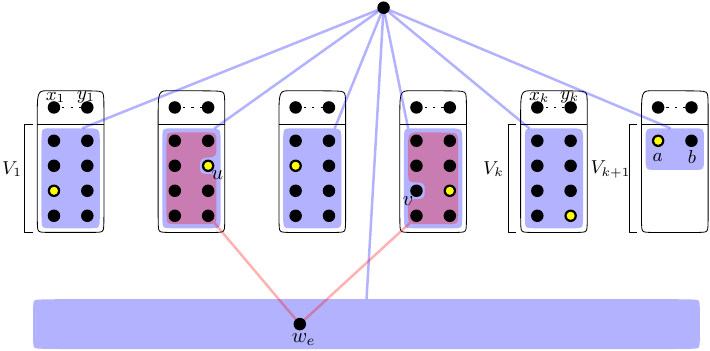}
        \caption{
        The graph $H$ in the reduction, adapted from the corresponding
        multicolored gadget in~\cite{DBLP:books/sp/CyganFKLMPPS15}.
        The set $V_i \cup \{x_i, y_i\}$ induces a clique minus the edge $\{x_i, y_i\}$. If edge $e$ connects $u \in V_i$ and $v \in V_j$, then we add a vertex $w_e$ adjacent to $(V_i \cup V_j) \setminus \{u, v\}$. We add a vertex $z$ adjacent to all vertices of $G'$ and every vertex $w_e$, and to none of the vertices $x_i, y_i$.}
        \label{fig:w1-hard}
    \end{figure}
    We have $k+1 \leq \gamma(H) \leq k+2$. 
    Due to space limitations, we omit the details, but one can show that $G'$ has an independent set containing exactly one vertex from each color class if and only if $\gamma(H)=k+1$, and that this holds if and only if $b(H)=1$.
    This completes the parameterized reduction.
\end{proof}

\section{Concluding Remarks}
Our results leave open the complexity of intermediate regimes between
fixed and unrestricted $d$, such as $d=O(\log n)$.
In particular, our classification contains no $\DeltaTwoP$-complete
regime, raising the question of whether some intermediate growth of $d$
yields $\DeltaTwoP$-completeness or a class characterized by limited
nondeterminism.

On the parameterized side, it remains open whether \bondage is
fixed-parameter tractable parameterized by $\operatorname{cw}(G)+d$.
For treewidth $k$, our algorithm has a doubly exponential dependence on
$k$. Such a dependence is known to be unavoidable under ETH for natural
problems at the second level of the polynomial hierarchy, including
$\exists\forall$-SAT~\cite{DBLP:conf/iwpec/LampisM17} and
3-Choosability~\cite{DBLP:conf/icalp/MarxM16}.
It is therefore natural to ask whether the same is true for \bondage.
Our present $\SigmaTwoP$-hardness reduction, however, is not designed
to preserve bounded treewidth and does not directly yield such a lower
bound; indeed, the constructed graph contains cliques of size linear in
the number of existential variables.
Thus, obtaining a $2^{2^{o(\mathrm{tw}(G))}}n^{O(1)}$ lower bound under
ETH, or a substantially faster treewidth-based algorithm, remains open.

\bibliographystyle{splncs04r}
\bibliography{main}

\end{document}